\documentclass[11pt]{article}

\usepackage[hidelinks]{hyperref}
\usepackage[table]{xcolor}
\usepackage{tabularray}
\usepackage[inline]{enumitem}
\usepackage{soul}
\usepackage{xcolor}
\soulregister\ref7
\soulregister\pageref7
\soulregister\cite7
\soulregister\eqref7
\soulregister\NCL7
\soulregister\AND7
\soulregister\OR7
\soulregister\NP7
\soulregister\PSPACE7
\soulregister\hanano7

\usepackage{booktabs,tabularx}

\usepackage{amsmath, amssymb, amsthm}
\usepackage[roman]{complexity}

\usepackage[a4paper, total={6in, 8in}]{geometry} 
\usepackage{fullpage}

\usepackage{algorithm,algpseudocode}

\usepackage{graphicx}
\graphicspath{ {./images/} }

\usepackage{subcaption}
\usepackage{float}

\usepackage{tikz}
\usetikzlibrary{positioning}
\tikzset{auto, node distance =2 cm and 3cm, on grid,
	semithick,
	dot/.style={circle,fill=black},
	state/.style ={circle, draw, black, text=black, minimum width =1cm},
	red/.style={ultra thick, color=red},
	blue/.style={ultra thick, color=blue}}

\NewTblrTheme{CaptionNotIndented}{%
     \DeclareTblrTemplate{caption-tag}{default}{}
     \DeclareTblrTemplate{caption-sep}{default}{}
     \DeclareTblrTemplate{caption-text}{default}{Table\hspace{0.25em}\thetable:\enskip\InsertTblrText{caption}}
}

\newtheorem{theorem}{Theorem}[section]
\newtheorem{proposition}[theorem]{Proposition}
\newtheorem{lemma}[theorem]{Lemma}
\newtheorem{example}[theorem]{Example}

\newtheorem{definition}[theorem]{Definition}

\newcommand{\hanano}{\ensuremath{\rm \textsc{Hanano}}}
\newcommand{\jelly}{\ensuremath{\rm \textsc{Jelly}}}

\newcommand{\NCL}{\ensuremath{\rm \textsc{NCL}}}
\newcommand{\AND}{\ensuremath{\rm \textsc{And}}}
\newcommand{\OR}{\ensuremath{\rm \textsc{Or}}}

\newcommand{\card}[1]{\lVert #1 \rVert}

\newcommand{\calD}{\ensuremath{{\cal D}}}
\newcommand{\calE}{\ensuremath{{\cal E}}}
\newcommand{\calF}{\ensuremath{{\cal F}}}
\newcommand{\calG}{\ensuremath{{\cal G}}}
\newcommand{\calJ}{\ensuremath{{\cal J}}}

\newcommand{\calL}{\ensuremath{{\cal L}}}
\newcommand{\calU}{\ensuremath{{\cal U}}}
\newcommand{\calV}{\ensuremath{{\cal V}}}

\newcommand{\LANGDEF}[3]{
\begin{center}
{\small 
\begin{tabularx}{0.98\columnwidth}{ll}
\toprule
\multicolumn{2}{c}{\textsc{#1}} \\
\midrule
{\bf Given:}   & \parbox[t]{0.75\columnwidth}{#2\vspace*{1mm}}  \\
{\bf Question:}& \parbox[t]{0.75\columnwidth}{#3\vspace*{.5mm}} \\ 
\bottomrule
\end{tabularx}
}
\end{center}
}

\title{Carrying is Hard: Exploring the Gap between Hardness for NP and PSPACE for the Hanano and Jelly no Puzzles}
\author{Michael C. Chavrimootoo and Jin Seok Youn\thanks{Current address: Uber Technologies, Inc., 1655 3rd St, San Francisco, CA 94158}\\Department of Computer Science\\Denison University\\Granville, OH 43023}
\date{August 20, 2026}

\newcommand{\RedBend}{\ensuremath{\rm \textsc{RedBend}}}
\newcommand{\BlueBend}{\ensuremath{\rm \textsc{BlueBend}}}

\begin{document}
\sloppy
\maketitle

\begin{abstract}
    The Hanano Puzzle is a one-player game with gravity, where the goal is to make colored blocks make contact with flowers of the corresponding color.  The game Jelly no Puzzle shares similar mechanics. In general, determining if a given level of each of the two games is solvable is \PSPACE-complete. There are also known restrictions under which determining if a level of Jelly no Puzzle is solvable is \NP-complete\@. We find that under the same restrictions, determining if a level of Hanano Puzzle is solvable remains \PSPACE-complete. We thus study several restrictions on Hanano, contrast them with known results about Jelly no Puzzle---at times giving membership in \P---and posit that the mechanism at the heart of the \PSPACE-hardness is the ability for blocks to carry other blocks.
\end{abstract}

\section{Introduction}

The complexity of games with movable blocks has been actively studied over the last three decades (see \cite{dem-hea:b:games-puzzles-comp} for a comprehensive historical overview, up to the time of writing), and much work has gone into devising frameworks to study those games~\cite{dem-hea:b:games-puzzles-comp,ani-bos-dem-dio-hen-lyn:c:door-pspace-hard,ani-chu-dem-dio-hen-lyn:c:checked-gadgets,ani-dem-hen-lyn:j:in-out-gadgets}, often viewing them as reachability problems. 
Games with gravity (i.e., games where an object falls if unsupported and there is no built-in mechanism to move the object up against the direction of gravity) often need some other considerations to utilize the strength of the aforementioned frameworks. Most recently, Chavrimootoo~\cite{cha:j:hanano} introduced the use of visibility representations to provide a more structured way of using the Nondeterministic Constraint Logic ($\NCL$) framework (defined later) to study sliding-block games with gravity, and they proved that determining if a level of the Hanano Puzzle is solvable is PSPACE-complete.

The Hanano Puzzle (Hanano, for short) is a one-player game on a 2D grid involving colored movable blocks and flowers that can be either red, blue, or yellow. Each movable colored block is marked on one of its four sides. A movable block can move to an adjacent (left/right) empty space in one step. When a colored movable block makes contact with a flower of the same color, a flower emerges from the marked side, thereby ``blooming'' the colored block. A colored block can only bloom once. If there are blocks in the way of the flower emerging, those blocks are pushed by the flower, as long as there is space to move them. A block blooming may result in a chain/sequence of blooms.
The goal of the game is to bloom every colored block in a given level. A level of Hanano can also contain movable gray blocks (which do not bloom) and immovable blocks (which can be viewed as walls).
The game involves gravity, so whenever a block is not supported (either by a movable or immovable block), it falls---which one can view as being part of the step in which the block was moved---until it is supported.
If a movable block $b$ is moved while it is supporting another movable block~$b'$, the supported block $b'$ does \emph{not} move with the supporting block $b$.
Flowers are fixed to blocks, and the only way to move a flower is to move the block it is attached to (thus flowers attached to immovable blocks can never be moved). 
We also mention that in Hanano, two movable blocks of width 1 can be swapped in one step as long as doing so does not require blocks to move to accommodate the new position of the taller block, i.e., the taller block fits in its new position. However, our constructions are not affected by the swap feature.

\begin{figure}
    \centering
    \includegraphics[width=0.35\linewidth]{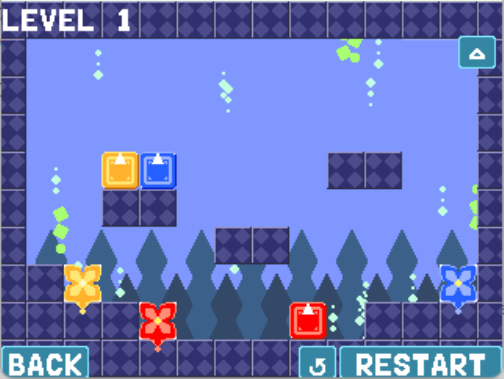}
    \caption{Level~1 of Hanano.}
    \label{fig:hanano-level1}
\end{figure}

Figure~\ref{fig:hanano-level1} shows three blocks of different colors, along with the corresponding flowers. Notice that the flowers are attached to wall-like structures, which correspond to  immovable gray blocks. Moreover, the arrowheads on the colored blocks indicate the aforementioned ``marks.''

\begin{figure}
    \centering

    \begin{subfigure}{0.32\linewidth}
        \centering
        \includegraphics[width=\linewidth]{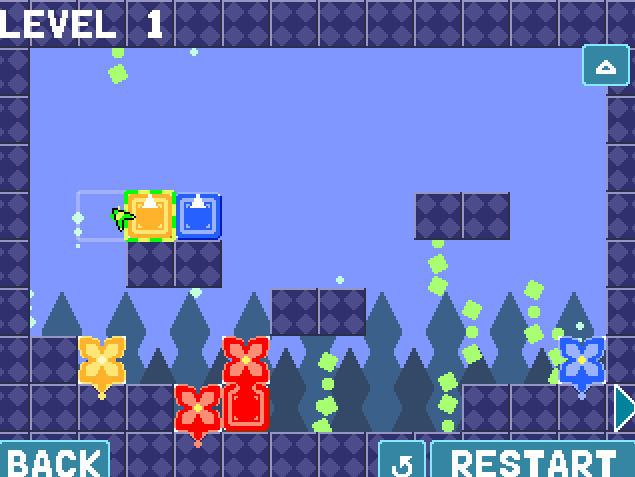}
        \caption{The red block moved to the left and bloomed.}
        \label{fig:hanano-moveleft}
    \end{subfigure}
    \hfill
    \begin{subfigure}{0.32\linewidth}
        \centering
        \includegraphics[width=\linewidth]{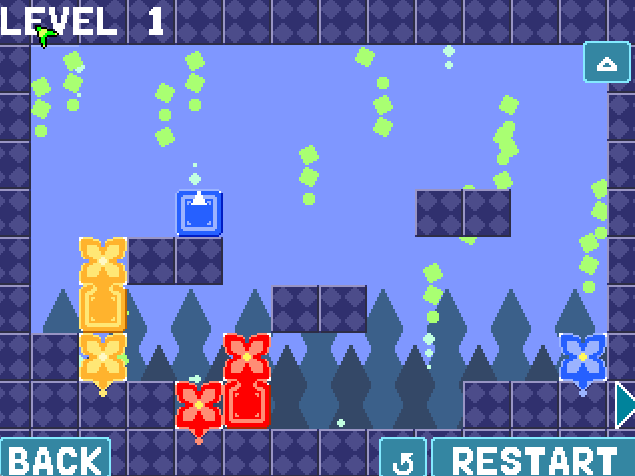}
        \caption{The yellow block moved left, fell, and bloomed.}
        \label{fig:hanano-bloom}
    \end{subfigure}
    \hfill
    \begin{subfigure}{0.32\linewidth}
        \centering
        \includegraphics[width=\linewidth]{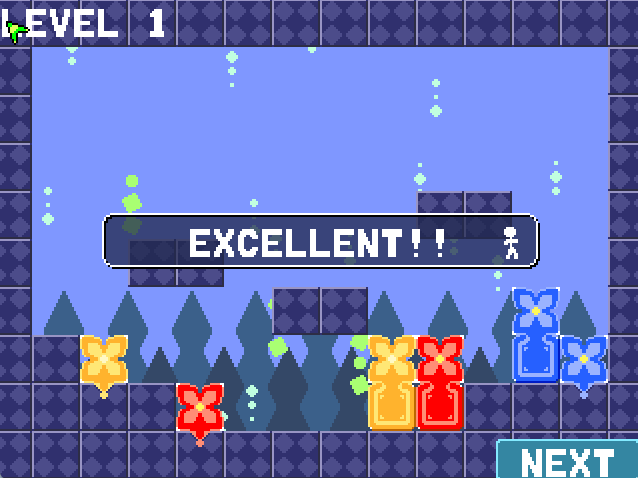}
        \caption{The level is completed/solved after every colored block blooms.}
        \label{fig:hanano-clear}
    \end{subfigure}
    \caption{Basic gameplay mechanics in Hanano.}%
    \label{fig:hanano-gameplay}
\end{figure}

Figure~\ref{fig:hanano-gameplay} illustrates the basic gameplay mechanics. As shown in Figure~\ref{fig:hanano-moveleft}, a movable block can be shifted to an empty adjacent space. As shown in Figure~\ref{fig:hanano-bloom}, if a block is no longer supported, it falls due to gravity until it is supported again. When a colored block contacts a flower of the same color, it blooms. Finally, as shown in Figure~\ref{fig:hanano-clear}, the level is cleared once every colored block has bloomed.

\begin{figure}[ht]
	\centering
	\begin{subfigure}[c]{0.4\linewidth}
		\centering
		\includegraphics[width=\linewidth]{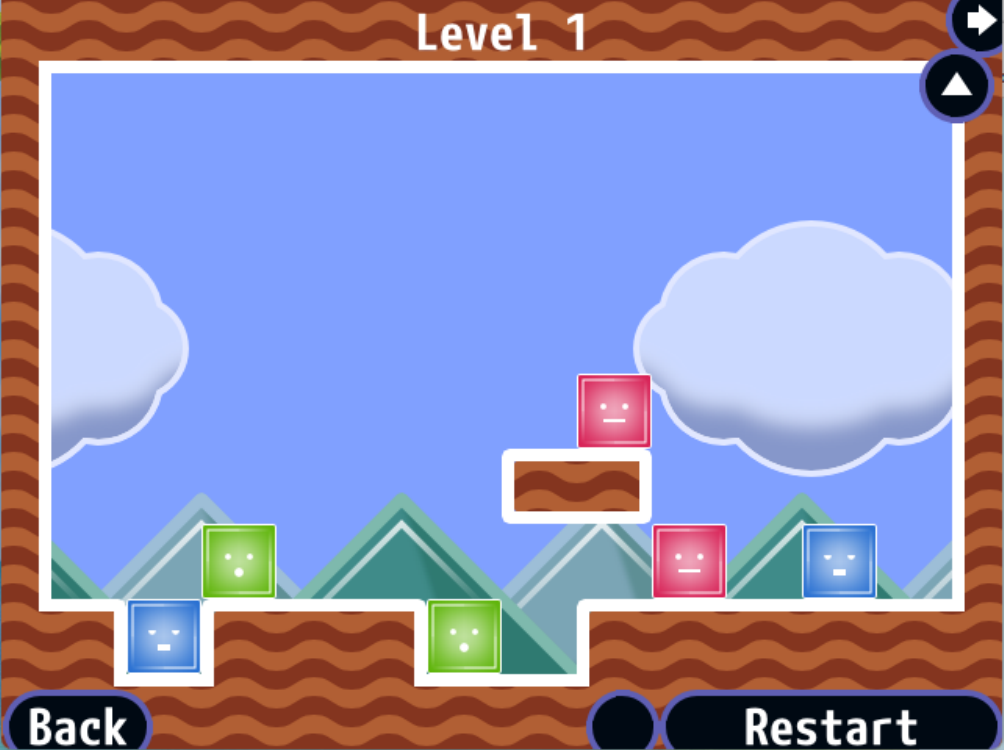}
		\caption{Initial configuration.}\label{fig:jelly-level1-before}
	\end{subfigure}
    \hspace*{2cm}
	\begin{subfigure}[c]{0.4\linewidth}
		\centering
		\includegraphics[width=\linewidth]{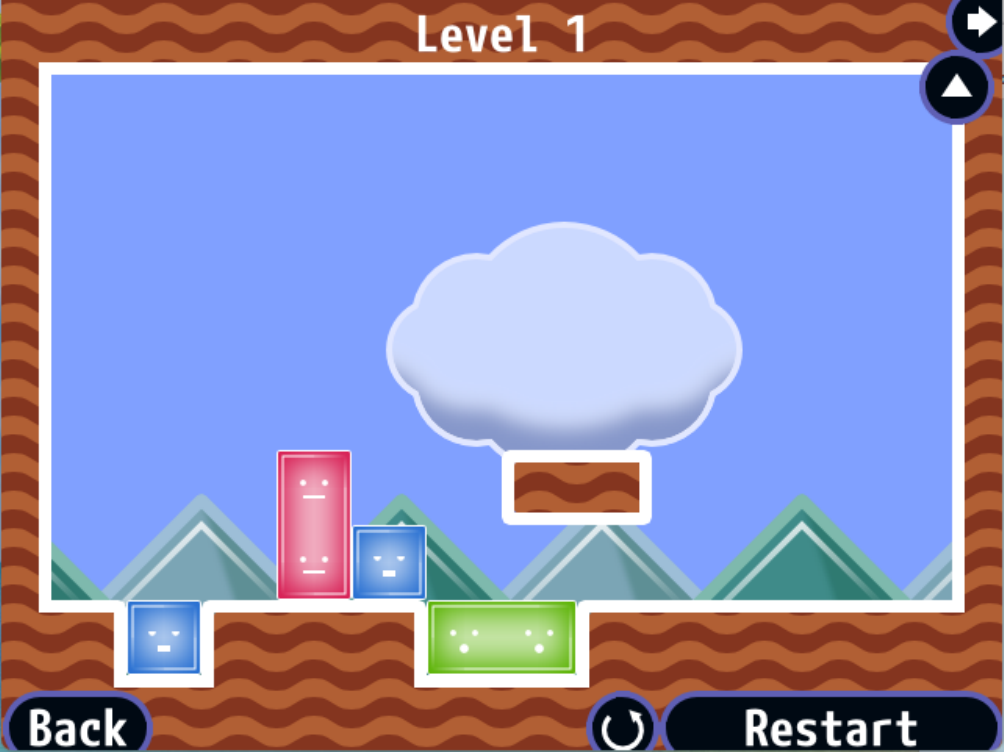}
		\caption{Example of merged jellies.}\label{fig:jelly-level1-merged}
	\end{subfigure}
    \caption{Two configurations of Level~1 of Jelly.}\label{fig:jelly-level1}
\end{figure}

Hanano is similar in its mechanics to Jelly no Puzzle~\cite{qro:url:jelly} (Jelly, for short).
A level of Jelly is very similar to a level of Hanano, except that instead of colored blocks and colored flowers, the level may contain colored ``jellies,'' which can be viewed as movable blocks.\footnote{Technically, what we call movable gray blocks are called black blocks/jellies in Jelly. However we use the gray color in our terminology to make the connection to Hanano clearer and avoid confusion within the paper.} When two jellies of the same color come into contact, they merge, thereby forming a new (possibly nonrectangular) jelly occupying the same space of as the original jellies. The goal of Jelly is to merge all the jellies of the same color. Figure~\ref{fig:jelly-level1} shows an example.

In Hanano, there is no merging of blocks, so contact with blocks that are already bloomed does not have any additional effect. In such cases, blocks remain ``independent'' and can be pushed or ``carried'' over one another. This enables certain movements and chain reactions that are not present in Jelly. 
Commensurate with prior work, we let \hanano\ denote the set of solvable levels of Hanano and we let \jelly\ denote the set of solvable levels of Jelly.
Liu and Yang~\cite{liu-yan:j:hanano} proved the NP-hardness of \hanano, and 
Chavrimootoo~\cite{cha:j:hanano} strengthened their result and proved that one-color $\hanano$ is \PSPACE-complete.  Crabtree and Mitsou~\cite{cra-mit:t:jelly} leveraged the main technique of Chavrimootoo~\cite{cha:j:hanano}---the use of visibility representations of planar $\NCL$ graphs---to prove the \PSPACE-completeness of \jelly\@. Interestingly, Yang~\cite{yan:c:jelly} showed that when restricted to one color and no movable gray blocks, \jelly\ is $\NP$-complete. We prove, in contrast, that \hanano\ remains \PSPACE-complete under that restriction.

\begin{figure}[h]
    \centering
    \includegraphics[width=0.45\linewidth]{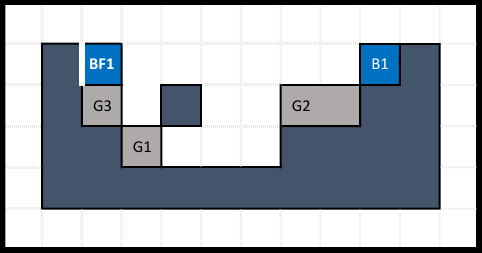}
    \caption{Carrying must be used to complete this level. The above paragraph provides the corresponding explanation.}\label{fig:width2-counterexample}
\end{figure}

In this paper, we consider several restrictions on $\hanano$ and $\jelly$ and demonstrate a hardness gap based on those restrictions. 
In particular, we
identify one game mechanism that seems to be the determiner for the jump from membership in $\NP$ to \PSPACE-completeness: carrying. 
Technically, Hanano and Jelly do not actually have a ``carrying'' feature, i.e., a feature that allows stacked blocks to move together by only moving the bottom-most block. However, one is able to simulate a similar action. 
Figure~\ref{fig:width2-counterexample}  shows an example of this. 
Assume the mark on B1 that indicates where the flower emerges from is at the top of B1\@.
To complete this level, one must make B1 move leftward to make contact with BF1\@.
The ``hard'' task here is to get B1 on top of the immovable block (in dark gray) two units to the right of G3, as once B1 is there, it suffices to slide G3 on top of G1 and slide B1 on top of G3 to complete the level.
One way to move B1 into the desired position is as follows: slide G1 three units right, slide B1 two units left, slide G2 one unit left, slide B1 one unit left, slide G2 one unit left (and notice at this point that G1 supports G2, so G2 does not fall), slide B1 two units left, and slide G1 three units left.
What this example captures is that G2 was effectively used to carry B1 from one point to another. Notice that B1 is not moved with G2. Rather, by gradually moving each block one at a time, the act of G2 carrying B1 was ``simulated.'' In this paper, when we speak of carrying, we really mean ``simulate carrying.''

Other work has explored the actions of games where an agent pushes blocks (see \cite{des-lia:t:push-1,mit-hardness:t:push-1} and the reference therein for more details)---possibly while the blocks are subject to the effects of gravity---%
but games like Hanano and Jelly are different because they do not involve an agent that is pushing blocks. In some sense, blocks in Hanano and Jelly ``move themselves,'' or simply, ``slide.''

\paragraph{Organization.} Section~\ref{s:prelim} defines necessary concepts surrounding $\NCL$, how it can be used to prove \PSPACE-hardness for games with gravity, and a formal model for games that we use in this paper. Section~\ref{s:pspace} proves that \hanano\ is PSPACE-complete for one color (even without gray blocks), and Section~\ref{s:carrying} proves additional results to support the conjectured ``dichotomy'' induced by the presence/absence of block carrying under $\hanano$ and $\jelly$, using the aforementioned formal model. Our discussion of relevant work is embedded within the most appropriate sections as opposed to being focused to a single section.

\section{Preliminaries}\label{s:prelim}

\subsection{Nondeterministic Constraint Logic}

The Nondeterministic Constraint Logic ($\NCL$) framework is a well-established framework for studying the complexity of games involving agents or blocks moving in a space~\cite{dem-hea:b:games-puzzles-comp}. An $\NCL$ graph is a directed graph with red edges and blue edges. The blue edges have weight 2 and the red edges have weight 1. $\NCL$ graphs have the additional constraint that for each vertex in the graph, the sum of weights of edges pointing to that vertex is at least 2. This constraint is also known as the inflow constraint. 
See Figure~\ref{f:ncl-example} for an example.

\begin{figure}[ht]
	\centering
	\begin{subfigure}[c]{0.49\linewidth}
		\centering
        \vspace*{-7mm}
		\begin{tikzpicture}[baseline,scale=0.7, transform shape,trim left=-10pt]
            \node[state] (A) {$A$};
            \node[state] (B) [right=of A] {$B$};
            \node[state] (C) [below =of B] {$C$};
            \node[state] (D) [below =of A] {$D$};
            \path[red, dashed, <-] (B) edge node [right, yshift=8pt] {\large1} (A);
            \path[blue, <-] (A) edge node [right, yshift=5pt] {\large2} (C);
            \path[blue, ->] (C) edge node [right] {\large2} (B);
            \path[blue, <-] (C) edge node {\large2} (D);
            \path[red, dashed, <-] (D) edge node [right] {\large1} (A);
            \path[red, dashed, <-] (D) edge[looseness=1.8,out=135]
             node[left, xshift=-10pt] {\large1} (B);
        \end{tikzpicture}
		
		\caption{An example of an $\NCL$ graph.}\label{f:ncl-example}
	\end{subfigure}
	\hfill
	\begin{subfigure}[c]{0.49\linewidth}
		\centering
		\begin{tikzpicture}[baseline,scale=0.35, transform shape]
    		\node [] (0) at (-8.5, 7) {};
    		\node [] (1) at (-10.25, 9) {};
    		\node [] (2) at (-8.5, 8) {};
    		\node [] (3) at (-10.25, 7) {};
    		\node [] (4) at (-5.25, 5) {};
    		\node [] (5) at (-8.5, 5) {};
    		\node [] (6) at (-5.25, 6) {};
    		\node [] (7) at (-3.5, 9) {};
    		\node [] (8) at (-3.5, 5.5) {};
    		\node [] (9) at (-5.25, 5.5) {};
    		\node [] (10) at (-5.25, 3.75) {};
    		\node [] (11) at (-10.25, 3.75) {};
    		\node [] (12) at (-5.25, 2.25) {};
    		\node [] (13) at (-3.5, 8.5) {};
    		\node [] (14) at (-10.25, 8.5) {};
    		\node [] (15) at (-3.5, 2.25) {};
    		\node [] (16) at (-10.25, 3) {};
    		\node [] (17) at (-3.5, 6.5) {};
    		\node [] (18) at (-8.5, 6.5) {};
    		\node [] (19) at (-8.5, 4.5) {};
    		\node [] (23) at (-7.5, 7.75) {\huge$A$};
    		\node [] (25) at (-3.75, 9.5) {\huge$B$};
    		\node [] (26) at (-6.25, 6) {\huge$C$};
    		\node [] (27) at (-10.25, 9.5) {\huge$D$};
    		\draw [<-,dashed] (3.center) to (0.center);
    		\draw [<-] (5.center) to (4.center);
    		\draw [->] (9.center) to (8.center);
    		\draw [->] (11.center) to (10.center);
    		\draw [<-,dashed] (14.center) to (13.center);
    		\draw [<-,dashed] (17.center) to (18.center);
    		\draw[ultra thick] (2.center) to (18.center);
    		\draw[ultra thick] (7.center) to (15.center);
    		\draw[ultra thick] (6.center) to (12.center);
    		\draw[ultra thick] (1.center) to (16.center);
    		\draw[ultra thick] (18.center) to (19.center);
        \end{tikzpicture}
		\vspace*{-7mm}
		\caption{A visibility representation of Figure~\ref{f:ncl-example}.}\label{f:visibility-rep-example}
	\end{subfigure}
    \caption{Example of an $\NCL$ graph and its visibility representation,
     reproduced from \cite{cha:j:hanano}. For  accessibility reasons, red edges are drawn with dashed lines and blue edges are drawn with solid lines.}\label{fig:ncl-representation}
\end{figure}

An edge in an $\NCL$ graph can be flipped (i.e., its direction can be reversed) if and only if doing so does not violate the inflow constraints on the graph.  For example, in Figure~\ref{f:ncl-example}, the edge $(C, A)$ cannot be flipped as doing so would violate the inflow constraint on the graph (as no edges would point to $A$), but the edge $(A, B)$ can be flipped as doing so does not violate the inflow constraints (as the blue edge $(C, B)$ would still be pointing to $B$).

\LANGDEF{Nondeterministic Constraint Logic (\NCL)}
{An $\NCL$ graph $G=(V, E)$ and a target edge $e \in E$.}
{Is there a sequence of edge flips on $G$ that eventually flips $e$?}

The \NCL\ problem is \PSPACE-complete even if~(1) $G$ is simple and planar, and~(2) if $V$ is restricted to $\OR$ vertices (vertices with exactly three incident blue edges) and $\AND$ vertices (vertices with exactly one incident blue edge and two incident red edges)~\cite{dem-hea:b:games-puzzles-comp}.
For example, in Figure~\ref{f:ncl-example}, $A$, $B$, and $D$ are $\AND$ vertices, while $C$ is an $\OR$ vertex.
For the rest of this paper, we assume the aforementioned restrictions on $\NCL$ graphs.

A visibility representation of a planar graph can be viewed as a planar drawing of the graph where the vertices are stretched out into columns or line segments so that all edges can be drawn as horizontal lines. (Note that the standard definition maps vertices to horizontal segments and edges to vertical lines, but the version we give---which is the one used in~\cite{cha:j:hanano,cra-mit:t:jelly}---is better suited for our purposes and is equivalent to the standard one in this setting.) Figure~\ref{f:visibility-rep-example} gives an example. A graph admits a visibility representation if and only if it is planar~\cite{tam-tol:j:visibility-representations}, so under our present assumptions, every $\NCL$ graph admits a visibility representation, and such a representation is polynomial-time---in fact, linear-time---computable~\cite{tam-tol:j:visibility-representations}.
Because Hanano and Jelly are both games with gravity, the process of giving reductions from NCL to such games is greatly simplified by using the visibility representations of NCL graphs~\cite{cha:j:hanano,cra-mit:t:jelly}, and we will explain that further in Section~\ref{sub:ncl-to-hanano}.

\subsection{Games and Configurations}

The definitions in this section are based on standard approaches to modeling games and their states (see for example, \cite[Chapter~5]{rus-nor:b:ai}).

\begin{definition}[Game]\label{def:game}
A game $G$ is a 3-tuple $(\calL, \sigma, \calF)$, where
\begin{enumerate*}[label={(\arabic*)}]%
    \item $\calL$ is a set of levels/instances of $G$,
    \item $\sigma : \calL \to 2^\calL$ is a successor function that when given a level $L \in \calL$ outputs a set of levels (those can be viewed as instances that can be reached in a single step), and 
    \item $\calF$ is a predicate over $\calL$ that indicates which instances are in a solved state.
\end{enumerate*}
\end{definition}

Let $G = (\calL, \sigma, \calF)$ be a game. 
The elements of $\calL$ capture each level and their possible states during gameplay.
A level $L'$ is said to be a configuration of a level $L$ if and only if there is a sequence $L_1, \ldots, L_k$ such that $L_1 = L$, for each $i \in \{2, \ldots, k\}$, $L_{i} \in \sigma(L_{i-1})$, and $L' = L_k$.
A level $L$ of a game $G$ is said to be solvable if there is a configuration $L'$ of $L$ such that $\calF(L')$ is true, and
we call such a configuration an accepting configuration of $L$\@.
Finally, $G$ has a naturally associated language $L_G$, which is the set of solvable levels of $G$\@. 

\begin{definition}[Configuration Graph]
    Let $L$ be an arbitrary level of a game $G = (\calL, \sigma, \calF)$. The \emph{configuration graph} of $L$ is a directed simple graph
    $\calG = (\calV, \calE)$  such that $\calV$ is the set of configurations of level $L$ and for each $L_1, L_2 \in \calV$, $(L_1, L_2) \in \calE$ if and only if $L_2 \in \sigma(L_1)$.
\end{definition}

Intuitively, we can interpret $\calG$ as a deterministic finite-state automaton with initial state $L$ and with the accepting configurations of $L$ representing the accepting states. Then each transition/edge represents a valid move (under the rules of $G$), and so a level being solvable under the rules of $G$ corresponds exactly to an ``accepting path'' in $\calG$. This allows us to view our problems as reachability problems. 
We formalize this intuition next.

\begin{lemma}\label{lemma:accepting-configuration}
    Let $G=(\calL, \sigma, \calF)$ be a game, let $L \in \calL$, and let $\calG$ be the configuration graph of $L$. It holds that $L \in L_G$ if and only if there is a path in $\calG$ from $L$ to an accepting configuration of $L$.
\end{lemma}
\begin{proof}
    If $L \in L_G$, then there is a sequence of moves from $L$ to an accepting configuration $L'$. Thus the aforementioned sequence of moves defines a path from $L$ to $L'$.
    On the other hand, if there is a path from $L$ to an accepting configuration $L'$, then the edges on the aforementioned path define the sequence of moves to go from configuration $L$ to $L'$. In other words, there is a sequence of moves to solve the level $L$, and so $L \in L_G$.
\end{proof}

It is easy to see that both Hanano and Jelly can be defined as 3-tuples in the above sense so that $\calF$ is polynomial-time computable and $\calL$ is polynomial-time decidable, i.e., $\calL \in \P$.

\section{One-Color PSPACE-Completeness}\label{s:pspace}

In this section, we prove that one-color \hanano\ remains \PSPACE-complete even with \emph{no movable gray blocks}, thus strengthening the \PSPACE-completeness result given by Chavrimootoo~\cite{cha:j:hanano} who proved that one-color \hanano\ is \PSPACE-complete with movable gray blocks. In fact, their proof uses nonrectangular movable blocks, and their paper asks for a construction that only uses rectangular movable blocks. Our construction fulfills this requirement.

\subsection{From NCL to Hanano}\label{sub:ncl-to-hanano}

Under $\NCL$, all moves are reversible in the sense that an edge flip on an edge~$e$ can be immediately undone by applying the edge flip operation on $e$ again. However, in Hanano the effects of gravity make it that certain moves cannot be undone in the same way. For example, moving the yellow block in Figure~\ref{fig:hanano-level1} to the left once cannot be undone by moving that block to the right once immediately after. Thus the effects of gravity must be handled with care in any reduction from $\NCL$\@. Chavrimootoo~\cite{cha:j:hanano} used the visibility representations of (planar) $\NCL$ graphs to guide their reduction to ensure that the moves in Hanano that correspond to $\NCL$ edge flips can be undone (in the above sense).

Their work establishes that only three types of gadgets need to be given: an $\OR$ gadget, an $\AND$ gadget, and a $\RedBend$ gadget. The $\AND$/$\OR$ gadgets satisfy the same constraints as the $\AND$/$\OR$ vertices (respectively). 
Let us now give some intuition for the $\RedBend$ gadget

Suppose we have an implementation of a gadget for vertex $D$ in Figure~\ref{f:visibility-rep-example}. That gadget would satisfy the same constraints as an $\NCL$ $\AND$ vertex, and a gadget for vertex $A$ should do the same. Because of the structure induced by the visibility representation, we may not be able to just reuse the same gadget. However, by including some mechanism to ``bend'' the topmost red edge of $D$ to be on the left side of $D$, we obtain a corresponding vertex that looks identical to $A$. This is essentially what the $\RedBend$ gadget enables us to do.
We now step back and describe how these gadgets are used in the reduction. Each vertex of the input \NCL\ graph is represented by a gadget, and they are connected by horizontal tunnels that correspond to the edges in the visibility representation. For each tunnel, there is a boundary block that is shared between the two endpoint gadgets. When this boundary block is inside one of the gadgets, it indicates that the corresponding edge is directed into the vertex represented by that gadget. Therefore, moving a shared boundary block from one gadget to the other simulates flipping the direction of the associated edge.
Each gadget is designed to satisfy the same constraints as the corresponding \NCL\ vertex. \AND/\OR\ gadgets are solvable only when the orientations of incident edges satisfy the corresponding inflow constraints. 
Violating the inflow constraints on a gadget during gameplay forces that gadget into an unsolvable state.
The $\RedBend$ gadget is a helper gadget used to reroute a block corresponding to a red edge to preserve the layout of the visibility representation.

\subsection{Main Result}

Before introducing our gadgets, we clarify the conventions used in our figures. Following the reduction of Chavrimootoo~\cite{cha:j:hanano}, $\AND$/$\OR$ gadgets represent the corresponding vertices in the visibility representation of the input \NCL\ graph, and $\RedBend$ gadgets are used to construct the $\AND$/$\OR$ gadgets. 

In our one-color construction, all colored blocks and flowers are blue, and every unbloomed colored block is set to bloom upward. One advantage of making colored blocks bloom upwards is that if a block blooms in a gadget, it cannot leave the gadget because it is too tall to fit in the tunnels. Blocks of size $2 \times 1$ or $1 \times 2$ represent pre-bloomed blocks. For these blocks, unlabeled sides indicate flower facets, while labeled sides act as passive facets that are used only for support or blocking. Under these conventions, a gadget is solvable if and only if the corresponding NCL constraints are satisfied.

\begin{theorem}\label{theorem:pspace-completeness}
    \emph{\hanano} remains $\PSPACE$-complete even when restricted to no movable gray blocks and single-colored blocks/flowers.
\end{theorem}

The rest of this subsection is devoted to proving Theorem~\ref{theorem:pspace-completeness}.

\begin{figure}[ht]%
    \centering
    \includegraphics[width=0.25\textwidth]{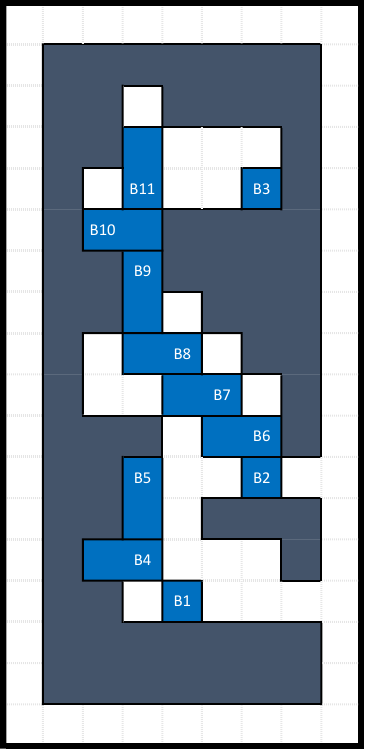}
    \caption{The $\RedBend$ gadget.}
    \label{fig:redbend}
\end{figure}

In Figure~\ref{fig:redbend}, B1 and B2 are the interface blocks corresponding to the two possible positions of the bent edge.

\begin{lemma}[One-color $\RedBend$]\label{lemma:red-bend}
The $\RedBend$ gadget in Figure~\ref{fig:redbend} is solvable if and only if either B1 or B2 blooms while supporting (either directly or indirectly) the spine (B6--B9).
\end{lemma}
\begin{proof}
First, the components of the gadget and the contact interaction rules are defined. In this gadget, blocks B4 to B11 are considered pre-bloomed blocks. The surfaces of these blocks are divided into two types according to their functionalities:

\begin{description}
    \item[Passive Facet:] The part marked with a label (e.g., ``B10''). This face functions as a simple physical wall and does not cause blooming even upon contact with unbloomed blocks.
    \item[Flower Facet:] The unlabeled part. This side performs the same function as a blue flower and triggers the blooming of an unbloomed block upon physical contact with this side.
\end{description}

For block B3 to bloom, it must come into contact with the flower facet of B10. In the initial configuration, B3 is separated from the flower facet of B10, and thus, contact cannot occur. Therefore, for effective contact to occur, B10 must be raised and aligned with B3. 

In this gadget, B10 is shifted up one unit through two routes:
\begin{enumerate}
    \item When B2 is supporting the spine (B6--B9) and B2 blooms.
    \item When B1 supports B4--B5, which in turn support the spine (B6--B9); when B1 blooms, it pushes up B4--B5 and lifts the entire spine.
\end{enumerate}

To support the spine, at least one of B1 or B2 must be in the gadget.
\begin{itemize}
    \item If B2 is in the gadget, B6--B9 is directly supported by B2. The B4--B5 structure can be stowed on the left side, and thus B1 is free to exit the gadget without causing the spine to collapse.
    \item Conversely, if B1 is in the gadget, B4--B5 supports the B6--B9. In this case, since B2's role becomes redundant, B2 is free to exit the gadget.
\end{itemize}

We now give our proof.

($\Longrightarrow$) Suppose the gadget is solvable. Then, B3 must come into contact with B10's flower facet. This is possible if and only if the spine (B6--B9) is raised by one unit. The spine is raised by one unit if and only if either B2 blooms or the B4--B5 structure is raised by one unit. The latter (uplift of B4--B5) occurs only when B1 blooms. In either case, the spine (B6--B9) must be supported by a bloomed B2, or by B4--B5 (which is in turn supported by a bloomed B1).

($\Longleftarrow$) Suppose that either B1 or B2 blooms while supporting the spine (B6--B9) either directly or indirectly. In either case, the spine rises by 1 unit (direct rise by B2 or indirect rise by B1 $\to$ B4--B5). The elevation of the spine shifts B10, exposing its flower facet and allowing B3 to make contact. Any other unbloomed block in the gadget can contact any flower facet of nearby bloomed blocks, thus making the gadget solvable.
\end{proof}

\medskip
\begin{figure}[ht]
    \centering
    \includegraphics[width=0.4\textwidth]{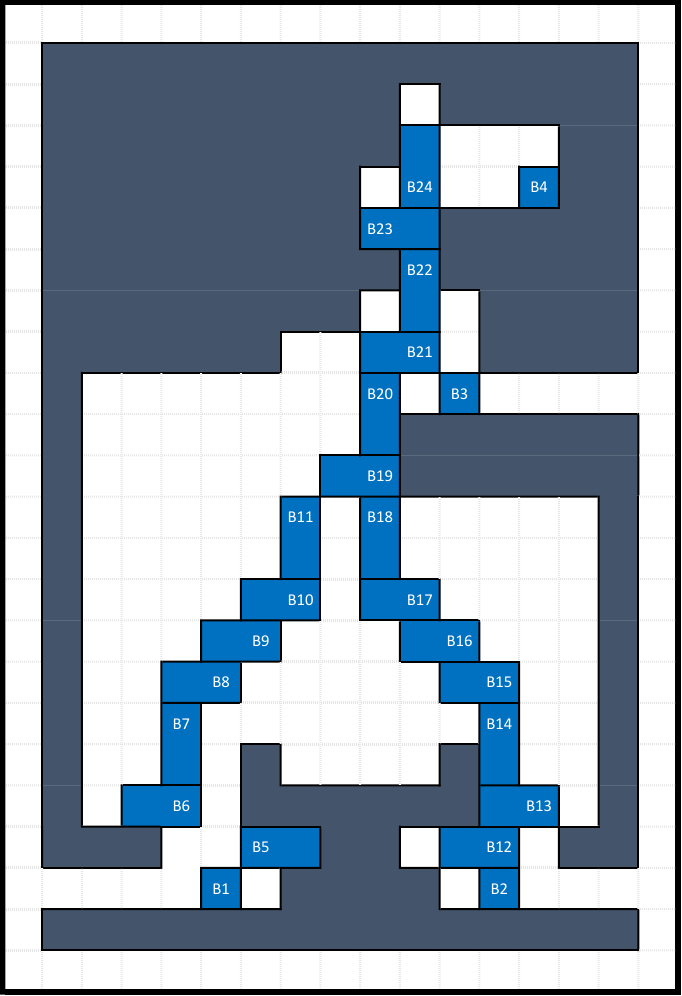}
    \caption{The $\OR$ gadget.}
    \label{fig:or}
\end{figure}

In Figure~\ref{fig:or}, the blocks B1, B2, and B3 represent the three incident edges of the corresponding \NCL\ \OR\ vertex.

\begin{lemma}[One-color $\OR$ Gadget]\label{lemma:or}
The gadget in Figure~\ref{fig:or} satisfies the same constraints as an $\NCL$ $\OR$ vertex.
\end{lemma}

\begin{proof}
First, the movable structures and configuration of the gadget are defined. The key to this gadget is controlling the position of the B21--B23, which is needed for the blooming of the output block B4. For this purpose, the internal structures are divided as follows:

\begin{description}
    \item[Left spine:] The B6--B11 + B19--B20 block, which can be stored on the left side.
    \item[Right spine:] The B13--B18 block, which can be stored on the right side.
    \item[Top Support:] The B21--B23 block, which can be directly supported by B3.
\end{description}

For block B4 to bloom, it must be in contact with the flower facet of B23. In the initial configuration, since B23 is located one unit below B4, the structure containing B23 must shift up for contact to occur.

The rise of B23 in this gadget is achieved through one of the following three routes:
\begin{itemize}
    \item Route 1: B3 blooms and shifts up B21.
    \item Route 2: B2 blooms and shifts up the entire structure of B12--B21.
    \item Route 3: B1 blooms and shifts up the entire structure of B5--B11 and B19--B21.
\end{itemize}

This design satisfies the inflow constraints for $\OR$ gadgets. That is, to position B23 in the shift-ready position, at least one of B1, B2, or B3 must exist in the gadget to provide support. Through the following state transition analysis, we show that no matter what input block enters, the support structure does not collapse, and the shift-ready position is maintained.

\begin{enumerate}
    \item If B3 is supporting: If B3 is in the gadget and supports B21, the remaining structures can be safely stowed. B19--B20 are stored on the left while being supported by the left spine (B6--B11), and the right spine (B13--B18), B5, and B12 are also located in their respective stowing positions. In this state, B1 and B2 are excluded from the supporting role, so they are free to exit.
    \begin{description}
        \item[Transition with B1:] If B1 enters, B1 moves B5 to support the left spine (B6--B11 + B19--B20) and moves it to the right as much as possible to extend the support range to B21. B3 and B2 are now free.
        \item[Transition with B2:] If B2 enters, it pushes B12 so that the right spine (B13--B18) shifts as far left as possible, allowing B19–B20 to be transferred onto (and supported by) the right spine. Once B19–B20 are in place, they support B21, and B3 becomes free.
    \end{description}

    \item If B1 is supporting: If B1 is in the gadget and supports the entire left spine.
    \begin{description}
        \item[Transition with B3:] If B3 enters, B3 supports B21. The remaining B19--B20 blocks are transferred to the left spine (B6--B11) and stowed by moving the spine to the left. After returning B5 to its original position, B1 is released from the load and becomes free.
        \item[Transition with B2:] If B2 enters, B2 uses B12 to move the right spine (B13--B18) to the left as much as possible. This takes over the load of B19--B21 supported by the left spine. After stowing the left spine (B6--B11) and B5 back to their original positions, B1 becomes free.
    \end{description}

    \item If B2 is supporting: If B2 is in the gadget and supports the right spine, including B19--B21.
    \begin{description}
        \item[Transition with B3:] If B3 enters, the same method as when B1 is supporting is applied. B3 supports B21, and the remaining B19--B20 are transferred to the left spine for stowing. Accordingly, B2 becomes free.
        \item[Transition with B1:] If B1 enters, B1 uses B5 to move the left spine (B6--B11) to the right as much as possible. Through this, the load of B19--B21 supported by the right spine is taken over. After stowing the right spine (B13--B18) and B12 back to their original positions, B2 becomes free.
    \end{description}
\end{enumerate}

During all transition processes, the structure remain supported at all times, which forcibly preserves the shift-ready position. We now give our proof.

($\Longrightarrow$) Suppose the gadget is solvable. If so, B4 must come into contact with the flower facet of B23. This is possible if and only if B23 rises by one unit. For B23 to rise, the substructure must be maintained in a shift-ready position without collapsing, which is only possible when one of the aforementioned routes (Routes 1, 2, and 3) is established. Therefore, at least one of B1, B2, or B3 must exist in the gadget to form the support base of the structures. If all inputs are absent, the structure drops, and blooming is impossible.

($\Longleftarrow$) Suppose at least one of B1, B2, or B3 blooms. Since the block currently maintains a shift-ready position in the gadget, it immediately shifts up the connected structure when it blooms.
\begin{itemize}
    \item If B3 Blooms $\to$ B21 rises.
    \item If B2 Blooms $\to$ B12--B21 rise.
    \item If B1 Blooms $\to$ B5--B11 and B19--B21 rise.
\end{itemize}
In any case, B23 rises by 1 unit, moves to the position of B4, and exposes the flower facet. Accordingly, since B4 can come into contact and bloom, the gadget is solvable.
\end{proof}

\begin{figure}[ht]
    \centering
    \includegraphics[width=0.3\textwidth]{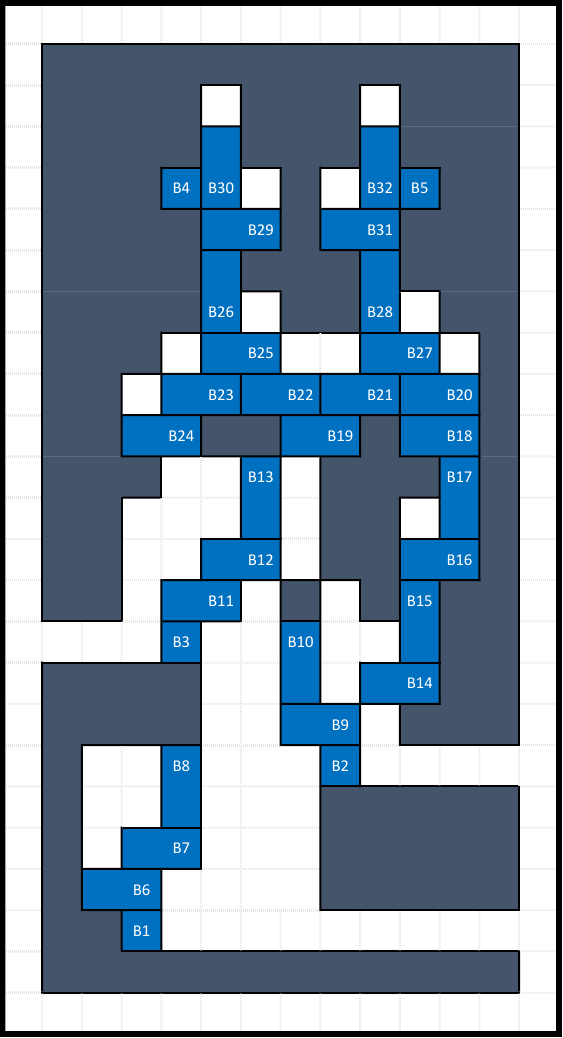}
    \caption{The $\AND$ gadget.}
    \label{fig:and}
\end{figure}

In Figure~\ref{fig:and}, B1 represents the blue edge and B2, B3 represent the two red edges of the corresponding \NCL\ \AND\ vertex.

\begin{lemma}[One-color $\AND$ Gadget]\label{lemma:and}
The gadget in Figure~\ref{fig:and} satisfies the same constraints as an $\NCL$ $\AND$ vertex, where B1 represents a blue edge and B2 and B3 represent red edges.
\end{lemma}

\begin{proof}
First, the Movable Structures and configuration of the gadget are defined. For this gadget to be solvable, both output blocks B4 and B5 must bloom. This requires the Upper Tree structure containing B29 and B31 to shift up by one unit to make contact with the flower facets of the output blocks. To control this, the internal structure is divided as follows:

\begin{description}
    \item[Left Spine:] The B11--B13 block group.
    \item[Right Spine:] The B9--B10 block group.
    \item[Central Spine:] The B6--B8 block group.
    \item[Upper Tree:] The complex structure comprising B14--B30, which interacts with the lower spines via B18, B19, and B24.
\end{description}

In this gadget, the simultaneous blooming of B4 and B5 is achieved through one of the following two actuation routes:

\begin{enumerate}
    \item Route 1 (Red/Red): B2 and B3 must bloom. B3 supports and lifts the Left Spine (B11--B13), and B2 supports and lifts the Right Spine (B9--B10) while the Right Spine remains in contact with B14 from below. The elevated Right Spine pushes up the right side (B14--B29) of the Upper Tree through B14, and the Left Spine pushes up the left side (B24--B30) of the Upper Tree by supporting the underside of B24. Together, these two forces raise both B29 and B31 by one unit.

    \item Route 2 (Blue): B1 blooms while holding up a stacked column. In this position, the Central Spine (B6--B8) sits on B1, the Right Spine (B9--B10) sits on the Central Spine, and the Left Spine (B11--B13) sits on top of the Right Spine. When B1 blooms, it pushes this whole stack upward. B13 pushes B19, lifting the Upper Tree and making both B4 and B5 bloom.
\end{enumerate}

This design satisfies the inflow constraints of the $\AND$ vertex. That is, to maintain the structure in the shift-ready position, either the support base {B1} or the combination {B2, B3} must exist. We show through the following state transition analysis that the supporting structure is preserved when valid inputs are present.

\begin{enumerate}
    \item If B1 is supporting: If B1 is in the gadget, B1 supports the Central Spine (B6--B8), the Right Spine (B9--B10) sits on top of it, and the Left Spine (B11--B13) sits on top of that. Since B1 bears the entire load, B2 and B3 are free to exit.
    \begin{description}
        \item[Transition with B2 \& B3:] For B1 to be free, both B2 and B3 must enter. Due to the immovable block located to the right of B11, B2 alone cannot support the Right Spine and the Left Spine at the same time. Therefore, B3 must enter and support the Left Spine (B11--B13), and B2 must enter and support the Right Spine (B9--B10). Once the load is distributed, B1 moves the Central Spine (B6--B8) to the left to be stowed. During this stowing process, the immovable block above B8 prevents accidental blooming. B1 then becomes free.
    \end{description}

    \item If B2 and B3 are supporting: B2 and B3 exist in the gadget, supporting the Right and Left Spines respectively in parallel. The Central Spine is stowed on the left.
    \begin{description}
        \item[Transition with B1:] B1 enters. B1 moves the Central Spine (B6--B8) stowed on the left to the center of the gadget. Then, the Right Spine (B9--B10) supported by B2 is transferred to the Central Spine, and then the Left Spine (B11--B13) supported by B3 is transferred to the Right Spine (Stacking). Since the entire structure is now supported by B1, both B2 and B3 become free.
    \end{description}
\end{enumerate}

Through this transition process, the shift-ready position is always preserved when the inflow is 2 or greater ({B1} or {B2, B3}). We now give our proof.

($\Longrightarrow$) Suppose the gadget is solvable. Then both B4 and B5 must bloom, which is possible if and only if the Upper Tree rises by one unit, exposing B29 and B31. To achieve this, the lower support structure must not collapse, so a combination satisfying the inflow constraints ({B1} or {B2, B3}) must exist in the gadget. If, for example, only B2 is present (inflow = 1), the structural constraint (immovable block) prevents it from supporting the Left Spine, causing the structure to drop and making blooming impossible.

($\Longleftarrow$) Suppose either B1 blooms or both B2 and B3 bloom.
\begin{itemize}
    \item If B1 blooms, Route 2 is activated: Central Spine $\to$ Right Spine $\to$ Left Spine $\to$ Upper Tree.
    \item If B2 and B3 bloom, Route 1 is activated: Parallel lifting of Left and Right Spines $\to$ Upper Tree.
\end{itemize}
In either case, B29 and B31 rise by one unit, allowing B4 and B5 to make contact and bloom. Thus, the gadget is solvable.
\end{proof}

\begin{proof}
\textbf{(of Theorem~\ref{theorem:pspace-completeness})}
    The \PSPACE\ membership follows by standard arguments, since a configuration of a Hanano level can be stored using polynomial space.

    For \PSPACE-hardness, we follow the reduction of Chavrimootoo~\cite{cha:j:hanano} from \NCL. For completeness, we provide the steps of the reduction here, following the steps of~\cite{cha:j:hanano}.
    
    Given an \NCL\ graph $G=(V, E)$ and a target edge $e=(u, v) \in E$, construct the visibility representation of $G$ in polynomial-time and construct an instance $H$ of \hanano\ based on the visibility representation, replacing each vertex with a corresponding gadget (\AND\ or \OR) and using tunnels to represent edges. Moreover, place the boundary blocks in the appropriate gadgets in a way that corresponds to the orientation of each edge.
    Notice that $H$ will be polynomially larger $(G, e)$.

    Let $b$ denote the boundary block that indicates the orientation of $e$. To ensure that the $H$ is solvable if and only if $b$ moves from the gadget for $v$ to the gadget for $u$ (without violating the inflow constraints), we ensure that every block in the gadget for $u$ can be bloomed if and only if $b$ is in the gadget for $u$ (without violating the inflow constraints).
    Let us explain how to do this by considering whether $u$ is an $\OR$ or $\AND$ vertex.%

    If $u$ is an $\OR$ vertex, then $b$ is either B1, B2, or B3 (using Figure~\ref{fig:or} as a reference). 
    Then we must find a way to prevent all other boundary blocks (that are not $b$) from being able to raise B19--B24 by one unit, all while requiring that those boundary blocks still have the ability to support B21--B24 from falling. In some sense, we are ``stunting'' (i.e., limiting its abilities) each boundary block; we will thus describe how to stunt the boundary blocks that are not $b$.
    We first consider the case where $b$ is B3. Then to stunt B1 and B2, it suffices to add an immovable gray block in the empty space above B21 as depicted in Figure~\ref{fig:stunt-or-b3} (to make it easier to see where new block is added, we've colored the block purple with a large, white-colored X).

    \begin{figure}[h]
        \centering
        \includegraphics[width=0.4\linewidth]{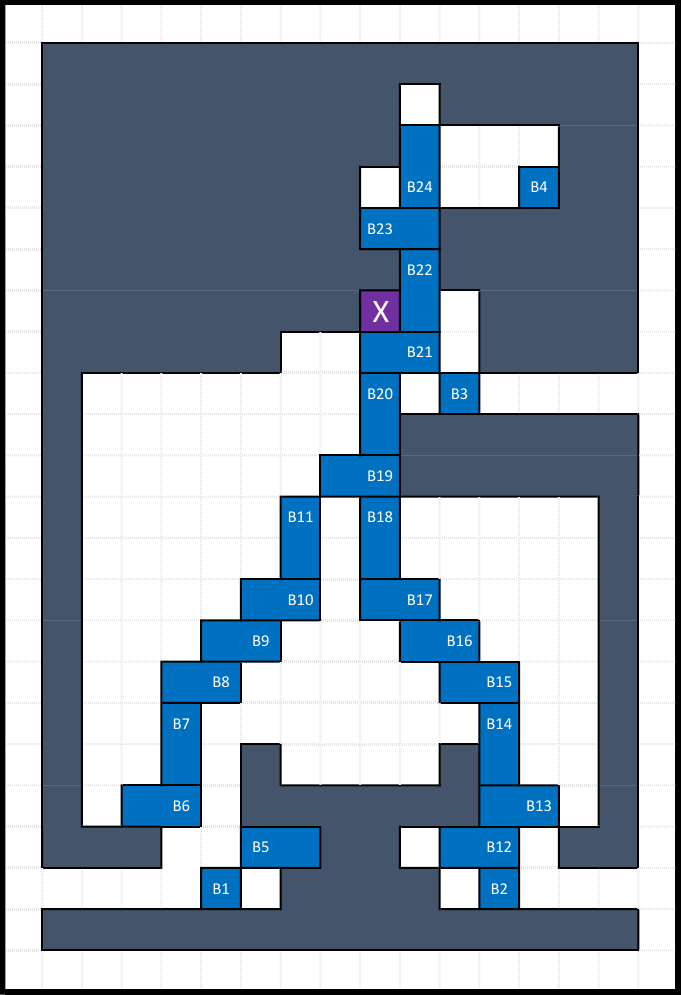}
        \caption{Stunted \OR\ gadget for when $b$ is B3.}
        \label{fig:stunt-or-b3}
    \end{figure}

    If $b$ is B1 or B2, the steps to stunt B3 seem to require nontrivial modifications to the gadget, so rather than presenting such new gadgets, we instead show how to construct them using the ones we already have. In Figure~\ref{fig:stunt-or-middle} we depict how to chain the $\RedBend$ gadget and the gadget in Figure~\ref{fig:stunt-or-b3} to yield an OR gadget that is, for our purposes, equivalent to one where B1 and B3 are stunted. 
    
\begin{figure}[htb!]
\centering
\begin{subfigure}[c]{0.4\linewidth}
\centering
\begin{tikzpicture}[scale=0.6, transform shape]
\node[rectangle, draw] (L) at (3, -1) [minimum height=5cm] {$\BlueBend$};
\node[rectangle, draw] (OR) at (6, -2) [minimum height=4cm] {Stunted \OR};
\node[rectangle, draw] (R) at (9, -4) [minimum height=3cm] {$\BlueBend$};

\draw[color=blue] (3.9, 1) -- node[below] {2} ++(7, 0);

\draw[color=blue,ultra thick] (7, -1) -- node[above] {2} ++(4, 0);
\draw[color=blue] (3.9, -3) -- node[above] {2} ++(1, 0);

\draw[color=blue] (7, -3) -- node[above] {2} ++(1, 0);

\draw[color=blue] (1.5, -5) -- node[above] {2} ++(6.5, 0);
\end{tikzpicture}
\caption{Constructing a stunted \OR\ gadget when $b$ is B2. The stunted \OR\ used in this construction is the one for the case where $b$ is B3.}\label{fig:stunt-or-middle}
\end{subfigure}
\hfill
\begin{subfigure}[c]{0.4\linewidth}
\centering
\begin{tikzpicture}[scale=0.6, transform shape]
\node[rectangle, draw] (L) at (3, -1) [minimum height=5cm] {$\BlueBend$};
\node[rectangle, draw] (OR) at (6, -2) [minimum height=4cm] {Stunted \OR};
\node[rectangle, draw] (R) at (9, -4) [minimum height=3cm] {$\BlueBend$};

\draw[color=blue] (3.9, 1) -- node[below] {2} ++(7, 0);

\draw[color=blue] (7, -1) -- node[above] {2} ++(4, 0);
\draw[color=blue] (3.9, -3.2) -- node[above] {2} ++(1, 0);

\draw[color=blue, ultra thick] (7, -2.8) -- node[above] {2} ++(1, 0);

\draw[color=blue, ultra thick] (1.5, -5) -- node[above] {2} ++(6.5, 0);
\end{tikzpicture}
\caption{Constructing a stunted \OR\ gadget when $b$ is B1. The stunted \OR\ used in this construction is the one for the case where $b$ is B2.}\label{fig:stunt-or-bottom}
\end{subfigure}
\caption{Schemas for constructing stunted $\OR$ gadgets.}
\end{figure}

In that figure, edges are undirected to represent that they can be in any orientation that satisfies the inflow constraints. Moreover, notice that the resulting gadget formed by making the connections indicated in that figure is effectively a stunted \OR\ gadget for the case where $b$ is B2. The edge corresponding to $b$ has been made thick. By our construction in Figure~\ref{fig:redbend}, the $\BlueBend$ gadget is the same as the $\RedBend$ gadget.
Similarly, Figure~\ref{fig:stunt-or-bottom} shows how to obtain a gadget that is, for our purposes, equivalent to an $\OR$ gadget where B2 and B3 are stunted. 

    If $u$ is an $\AND$ vertex, then $b$ is either B1, B2, or B3 (using Figure~\ref{fig:and} as a reference). 
    Just as we did in the previous paragraph, we now describe how to stunt each boundary block. 
    If $b$ is B1, then both B2 and B3 must be stunted. To do so, place a $1 \times 1$ immovable gray block in the empty square immediately above B14 (to stunt B2) and a similar immovable block in the corner left of B13's flower facet (to stunt B3) as depicted in Figure~\ref{fig:stunt-and-b2b3}.

    \begin{figure}[h]
        \centering
        \begin{subfigure}[c]{0.35\linewidth}
            \centering
            \includegraphics[width=\linewidth]{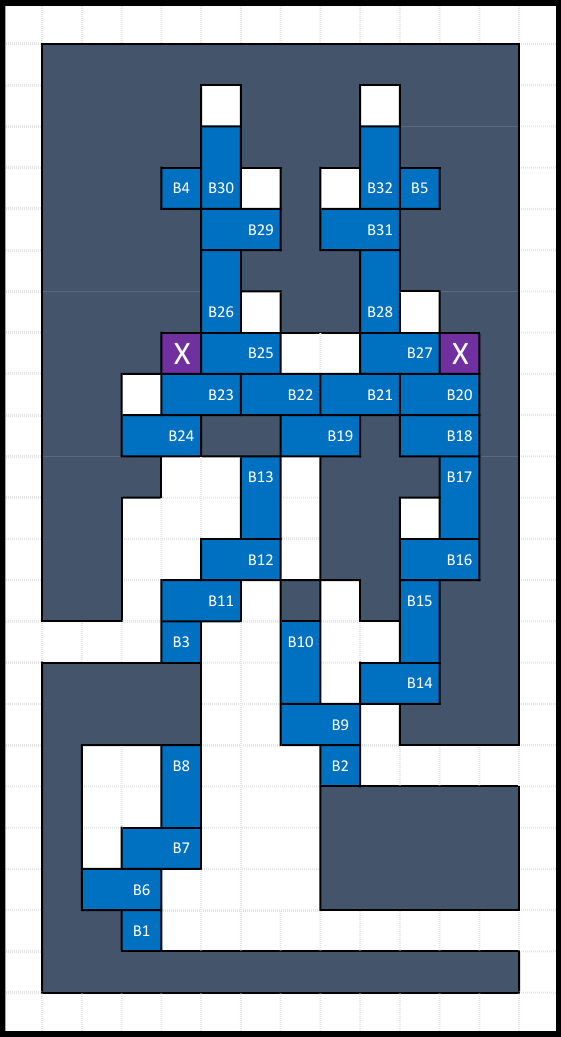}
            \caption{Stunted \AND\ gadget for when $b$ is B1.}
            \label{fig:stunt-and-b2b3}
        \end{subfigure}
        \hfill
        \begin{subfigure}[c]{0.35\linewidth}
            \centering
            \includegraphics[width=\linewidth]{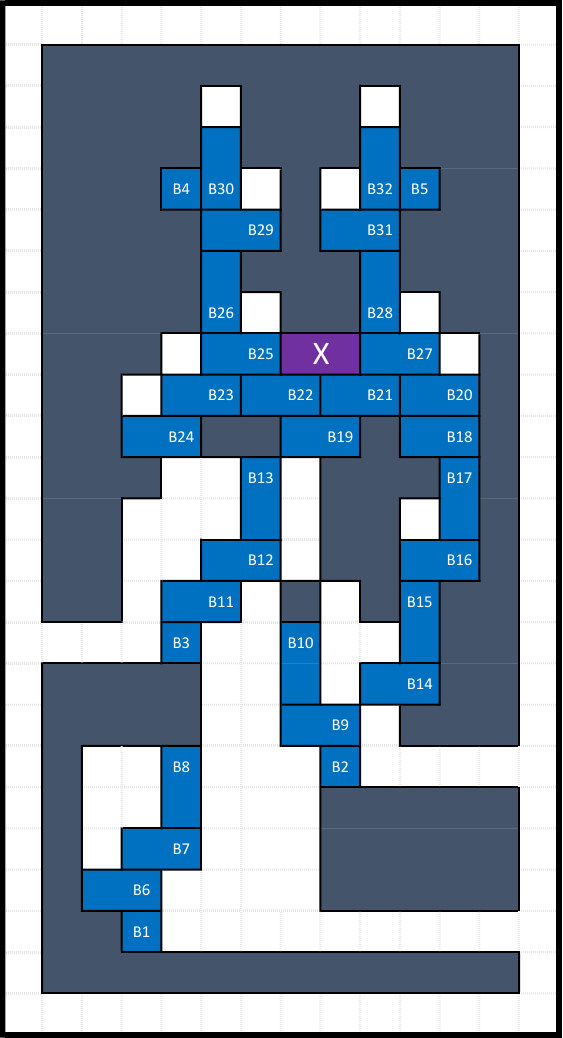}
            \caption{Stunted \AND\ gadget for when $b$ is B2 or B3.}
            \label{fig:stunt-and-b1}
        \end{subfigure}
        \caption{Stunted \AND\ gagets.}
        \label{fig:stunted-and-gadgets}
    \end{figure}

    However, if $b$ is either B2 or B3, then only B1 should be stunted. To do so, place a $1 \times 2$ immovable gray block in the empty space immediately above B21 and B22 as depicted in Figure~\ref{fig:stunt-and-b1}.

    Since each edge flip in $G$ corresponds to a traversal between gadgets of a boundary block, $(G, e) \in \NCL$ if and only if $H \in \hanano$.
    Finally, the above-described reduction is polynomial-time computable, so \hanano\ remains \PSPACE-complete under those restrictions.
\end{proof}

Figure~\ref{fig:reduction-sketch} provides an example of how to assign corresponding gadgets to the vertices of a \NCL\ graph's visibility representation by using the example in Figure~\ref{fig:ncl-representation}. Readers may notice that the structure of the gadgets we give do not correspond exactly to the structure of the gadgets needed here. For example, the gadget for $D$ has all its tunnels on the same side, which is not the same structure as what our \AND\ gadget has. However, \cite{cha:j:hanano} details how to handle such cases (essentially by showing that by implementing a version of an \OR, \AND, and \RedBend\ gadget, all the variations one may need can be constructed by chaining the existing gadgets), and we refer interested readers to that paper for the details of the proof.

\begin{figure}[h]
\centering
\begin{tikzpicture}[scale=0.7, transform shape]
\node[rectangle, draw] (D) at (0, 0) [minimum height=7cm] {\AND\ ($D$)};
\node[rectangle, draw] (A) at (3, -1) [minimum height=3cm] {\AND\ ($A$)};
\node[rectangle, draw] (C) at (6, -2) [minimum height=3cm] {\OR\ ($C$)};
\node[rectangle, draw] (B) at (9, 0) [minimum height=6cm] {\AND\ ($B$)};

\draw[color=red, <-, dashed] (0.9, 2) -- node[above] {1} ++(7.2, 0);
\draw[color=blue, ->] (0.9, -3) -- node[above] {2} ++(4.3, 0);
\draw[color=blue, ->] (6.8, -1) -- node[above] {2} ++(1.3, 0);
\draw[color=blue, <-] (3.9, -1.5) -- node[above] {2} ++(1.3, 0);
\draw[color=red, <-, dashed] (0.9, -2) -- node[above] {1} ++(1.1, 0);
\draw[color=red, ->, dashed] (4, 0) -- node[below] {1} ++(4.1, 0);
\end{tikzpicture}
\caption{An example showing how the \hanano\ gadgets map to vertices in the visibility representation.}\label{fig:reduction-sketch}
\end{figure}

\section{Carrying is Hard}\label{s:carrying}

We first provide the following result to support our intuition that carrying is hard by considering the restriction on levels of Hanano where movable blocks always have width 1 and thus cannot carry other movable blocks.

\begin{theorem}\label{theorem:width-1-hanano}
    When restricted to movable blocks of width 1 (including post-bloom), \emph{\hanano} is $\NP$-complete.
\end{theorem}
\begin{proof}
        The $\NP$-hardness was given by Liu and Yang~\cite{liu-yan:j:hanano}. We focus on $\NP$ membership here.
        We appeal to the notion of ``switchbacks'' used by Yang~\cite{yan:c:jelly} to prove that 1-color \jelly\ is in \NP\ and by Espasa et al.~\cite{esp-gen-mig-nig-sal-vil:c:puzznic} to prove that Cubic (an $\NP$-hard game with movable games and gravity~\cite{fri:a:cubic}) is in $\NP$.

        Let $H$ be a level of Hanano, set on an $n \times m$ grid, subject to the theorem's restrictions.
        Let $C$ be a configuration of $H$, and 
        number the rows of $C$ as $1, \ldots, n$, starting from the bottom.
        Let $n(C, i)$ denote the number of cells in row $i$ of $C$ containing a movable element. We define a fall metric on $C$ as $f(C) = \sum_{i = 1}^{n} i \times n(C, i)$. 
        \begin{example}
            Let $C$ be the configuration of the level of Hanano in Figure~\ref{fig:hanano-level1}. Then $n(C, 2) = 1$, $n(C, 6) = 2$, and for all other $i$, $n(C, i) = 0$, so $f(C) = (1 \times 2) + (2 \times 6) = 14$.
        \end{example}

        We now prove that the fall metric is bounded by a polynomial and also that for any configuration, there are only a polynomial number of distinct realizable configurations without changing the fall metric. It is also important to note that the fall metric is not nonincreasing, as a blooming block can lead to a new configuration with a higher fall metric. However, the number of times the fall metric can increase is bounded by a polynomial, namely $nm$, which is the maximum number of colored blocks.

        \begin{lemma}\label{nested-lemma:configurations}
            Let $C$ be a configuration of $H$, which is set on an $n \times m$ grid. Then the number of distinct configurations $C'$ that can be reached from $C$ without changing the fall metric is at most $nm^2$.
            Moreover, $f(C) \leq mn(n+1)/2$.
        \end{lemma}
        \begin{proof}
            Let $C'$ be a configuration that can be reached from $C$ by moving blocks without ever changing the fall metric, i.e., no blooms or drops occurred when going from $C$ to $C'$. Suppose a block $b$ is moved in one direction and then returns to its original location when going from $C$ to $C'$. Because blocks have width~1, no carrying can occur, so the foregoing sequence of moves made by $b$ such that $b$ started in one position and returned to the same position can be eliminated. Thus $b$ can only move in one direction before a change in the fall metric must occur. So $b$ can move at most $m$ positions. Moreover, there are at most $nm$ movable blocks, so there are at most $nm^2$ configurations that can be reached from $C$ without changing the fall metric.
            Also, notice that for each $i$, $n(C, i) \leq m$, as that is the number of cells in that row. So
            $f(C) = \sum_{i=1}^n \left(i \times n(C, i)\right) \leq \sum_{i=1}^n (i \times m) = m \left(\sum_{i=1}^n i\right) = mn(n+1)/2.$\hfill(Lemma~\ref{nested-lemma:configurations})
        \end{proof}

        Finally, taking stock of all our given bounds, if $H$ is solvable, there is a solution that makes at most 
        \begin{align*}
            \underset{\substack{\text{number of increases}\\\text{in the fall metric}}}{nm} \times 
            \underset{\text{bound on the fall metric}}{mn(n+1)/2} \times
            \underset{\substack{\text{moves without}\\\text{changing fall metric}}}{nm^2}
        \end{align*}

        moves, which is polynomial in $n$ and $m$. %
\end{proof}

To contrast the above result, our proof of Theorem~\ref{theorem:pspace-completeness} has no movable gray blocks, but has blue blocks of width 2, which allows for carrying to occur. Moreover, the argument crucially relies on blocks having width 1; Figure~\ref{fig:width2-counterexample} demonstrates a level of Hanano with only one width-2 block for which the argument of Theorem~\ref{theorem:width-1-hanano} fails.
This also highlights an interesting difference with new results on deterministic games, such as Push-1~\cite{des-lia:t:push-1,mit-hardness:t:push-1} which was recently shown to be \PSPACE-complete using gadgets with only width-1 blocks.

We will now shift our attention to a simple restriction on Hanano and Jelly\@. Namely, we will allow for arbitrary sizes on movable blocks---thereby allowing for carrying to happen---but we will restrict the number of such blocks to a constant. In doing so, we will observe a dramatic drop in the complexity of Hanano and Jelly, going from $\PSPACE$-complete to being in $\P$\@. 
The real utility of configurations in this paper is captured here. %

\begin{lemma}\label{lemma:easy-graph}
    Let $G = (\calL, \sigma, \calF)$ be a game. If 
    \begin{enumerate*}[label={(\arabic*)}]
        \item $\sigma$ is polynomial-time computable,
        \item the size of the configuration graph of each level $L$ is polynomial in $|L|$, and
        \item $\cal F$ is polynomial-time computable, 
    \end{enumerate*}
    then $L_G \in \P$.
\end{lemma}
\begin{proof}
    Let $L$ be an arbitrary level of $G$. Our polynomial-time algorithm for $L_G$ is the following, which is based on the  Breadth-First Search (BFS) algorithm.
    \begin{algorithmic}[1]
        \Procedure{Decide-$L_G$}{$L$}
            \State Let $S \gets \{L\}$
            \State Let $Q$ be an empty queue
            \State $Q$.\Call{Enqueue}{$L$}
            \While{$Q$ is not empty}
                \State $L' \gets Q.$\Call{Dequeue}{\,}
                \ForAll{${\hat L} \in \sigma(L')$} \Comment{Polynomial-time computable by condition~(1)}
                    \If{$\calF(\hat L)$} \Comment{Polynomial-time computable by condition~(3)}
                        \State Halt and accept
                    \ElsIf{$\hat L \not\in S$}
                        \State $Q$.\Call{Enqueue}{$\hat L$}
                        \State $S \gets S \cup \{\hat L\}$
                    \EndIf
                \EndFor
            \EndWhile
            \State Halt and reject
        \EndProcedure
    \end{algorithmic}

    By Lemma~\ref{lemma:accepting-configuration}, the correctness of the above algorithm follows. Moreover, by condition~(2), the configuration graph $\calG = (\calV, \calE)$ of $G$ is polynomial-time computable, so $\card{\calV}$ is polynomial in $|L|$. Since $\card{S} \leq \card{\calV}$, the while-loop iterates at most $\card{\calV}$ times, so the algorithm runs in polynomial time.
\end{proof}
In the above lemma, one may be tempted to argue that the result that ``$L_G \in \P$'' follows directly from condition~(2), but that may not be true. 
Certainly the need for condition~(2) is intuitive as games like Hanano and Jelly do not satisfy condition~(2), but they do satisfy conditions~(1) and~(3), and so under the widely-believed complexity-theoretic assumption that $\P \neq \PSPACE$, $\hanano$ and $\jelly$ are not in $\P$.
To further motivate the other two conditions, we prove that if one does not require either condition~(1) or condition~(3), then one can design a game $G$ that satisfies the other two conditions and yet $L_G \in \P$.

\begin{proposition}\label{prop:necessary-conditions}
     For each $k \in \{1, 3\}$, there is a game $G = (\calL, \sigma, \calF)$ that only fails to satisfy condition~($k$) of Lemma~\ref{lemma:easy-graph} (while satisfying the other two conditions of that lemma) and yet $L_G \not\in \P$.
\end{proposition}
\begin{proof}
Let $Z$ be an undecidable language over a unary alphabet $\Gamma$.
\noindent\textbf{Case for \boldmath$k=1$.}
We will define a game based on $Z$ that satisfies conditions~(2) and~(3), but not condition~(1).
For a given $x \in \Gamma^*$, define the graph $\calG_x = (\calV_x, \calE_x)$, where $\calV_x = \{v_0, v_1, \ldots, v_{|x|+1}\}$ and 
$E_x = \emptyset$.

Now define the game $G_1 = (\calL_1, \sigma_1, \calF_1)$ as follows:
\begin{enumerate}
    \item $\calL_1 = \{(\calG_x, v_i) \mid x \in \Gamma^* \land v_i \in \calV_x\}$,
    \item $\sigma_1((\calG_x, v_i)) = 
        \{v_{i+1} 
        \mid x \in Z \land 0\leq i < |x|+1\}$, and
    \item $\calF_1((\calG_x, v_i))$ is true if and only if $v_i = v_{|x|+1}$.
\end{enumerate}

Intuitively, the game is one of traversing a graph from one point to another. Each level of $G_1$ is a pair consisting of a graph and the current location on the graph. However, the ``hardness'' of the game is determining what the next move is. Indeed, for an arbitrary $x \in \Gamma^*$, $(\calG_x, v_0)$ is a solvable level of $G_1$ if and only if $x \in Z$, 
so if $L_{G_1} \in \P$, then $Z \in \P$, which is impossible.
It can be easily verified that conditions~(2) and~(3) are satisfied. Therefore condition~(1) is not satisfied.
\smallskip

\noindent\textbf{Case for \boldmath$k=3$.}
We define a game similar to $G_1$ where the ``hardness'' of the game is instead ``encoded'' into the predicate to determine if a final configuration was reached.
For a given $x \in \Gamma^*$, define the graph $\calJ_x = (\calU_x, \calD_x)$, where $\calU_x = \{u_0, u_1, \ldots, u_{|x|+1}\}$ and 
$\calD_x = \{(u_i, u_{i+1}) \mid 0 \leq i \leq |x|\}$.

Now define the game $G_3 = (\calL_3, \sigma_3, \calF_3)$ as follows:
\begin{enumerate}
    \item $\calL_3 = \{(\calJ_x, u_i) \mid x \in \Gamma^* \land u_i \in \calU_x\}$,
    \item $\sigma_3((\calJ_x, u_i)) = \{v \in \calU_x \mid (u_i, v) \in \calD_x\}$, and
    \item $\calF_1((\calG_x, v_i))$ is true if and only if $v_i = v_{|x|+1}$.
\end{enumerate}

Similarly to the analysis for $G_1$, for an arbitrary $x \in \Gamma^*$, $(\calJ_x, u_0)$ is a solvable level of $G_3$ if and only if $x \in Z$, 
so if $L_{G_3} \in \P$, then $Z \in \P$, which is impossible.
It can be easily verified that conditions~(1) and~(2) are satisfied. Therefore condition~(3) is not satisfied.
\end{proof}

While the constructions in the proof of Proposition~\ref{prop:necessary-conditions} are \emph{not natural}, i.e., they do not depict real-world games, the proposition highlights that simply arguing that a game is polynomial-time solvable because ``each level has a polynomial number of configurations'' is not a valid claim. 

We find the next result also contributes to the ``carrying'' narrative as it highlights that a constant number of blocks involved in carrying is not enough for computational hardness. 

\begin{theorem}\label{theorem:hanano-constant-movable}
    For each $k \geq 0$,
    \emph{$\hanano$} is in $\P$ when restricted to $k$ movable blocks.
\end{theorem}
\begin{proof}
    Let $(\calL, \sigma, \calF)$ be a 3-tuple defining Hanano; under the rules of Hanano both $\sigma$ and $\calF$ are polynomial-time computable.
    Let $k$ be a nonnegative integer and consider an arbitrary element $H \in \hanano$ with at most $k$ movable blocks laid out on an $n \times m$ grid. 
    We assume that $k > 0$, as otherwise the instance is trivial to solve. 
    Let $\calG = (\calV, \calE)$ be the configuration graph of $H$\@.
    Since both $\sigma$ and $\calF$ are polynomial-time computable, in
     light of Lemma~\ref{lemma:easy-graph}, it suffices to show that the size of $\calG$ is bounded by a polynomial in $|H|$\@.
    Since $\calG$ is a simple graph, it suffices to prove that $\card{\calV}$ is bounded by a polynomial in $|H|$. 
    Movable blocks can be either colored blocks, which can be bloomed or unbloomed, or movable gray blocks. 
    Let us now consider the number of distinct configurations of that level. 
    Since colored blocks can be bloomed or unbloomed, there are at most $2^k$ ``states'' for the $k$ movable blocks. For a given state of the movable blocks, each such block has its bottom left corner in at most one of $nm$ positions, so the number of possible arrangements given a particular state for the $k$ movable blocks is at most $(nm)^{k}$. Therefore the number of distinct configurations of $H$ is at most $2^k(nm)^k$, which is polynomial in $|H|$ as $k$ is a constant and $nm \leq |H|$.
\end{proof}

The analogous result also holds under $\jelly.$

\begin{theorem}\label{theorem:jelly-constant-movable}
    For each $k \geq 0$,
    \emph{$\jelly$} is in $\P$ when restricted to $k$ movable blocks.
\end{theorem}
\begin{proof}
    Let $(\calL, \sigma, \calF)$ be a 3-tuple defining Jelly; under the rules of Jelly both $\sigma$ and $\calF$ are polynomial-time computable.
    Let $k$ be a nonnegative integer and consider an arbitrary element $J \in \jelly$ with at most $k$ movable blocks laid out on an $n \times m$ grid. 
    We assume that $k > 0$, as otherwise the instance is trivial to solve. 
    Let $\calG = (\calV, \calE)$ be the configuration graph of $J$\@.
    Since both $\sigma$ and $\calF$ are polynomial-time computable, in
     light of Lemma~\ref{lemma:easy-graph}, it suffices to show that the size of $\calG$ is bounded by a polynomial in $|J|$\@.

    Since $\calG$ is a simple graph, it suffices to prove that $\card{\calV}$ is bounded by a polynomial in $|J|$. 
    Movable blocks can be either colored blocks (i.e., jellies) or movable gray blocks. 
    Whether jellies are merged (or not) does not need to be considered here for this upper bound.
    Each movable block can be in one of $nm$ positions, so the number of distinct configurations is at most $(nm)^k$, which is polynomial in the size of $J$ as $k$ is a constant. 
\end{proof}
\section{Summary and Future Work}

We have strengthened a prior result~\cite{cha:j:hanano} on the complexity of one-color $\hanano$ by proving that it remains $\PSPACE$-complete with no movable gray blocks. 
We identified restrictions on $\hanano$ that ``drop'' its complexity.
We argued that ``carrying'' is the main mechanism that induces hardness here, and we used results about $\hanano$ and $\jelly$ to support that argument. The comparison is summarized in Table~\ref{table:summary}. 
We provide below a few directions for future work.
\begin{enumerate}[wide,nosep]
    \item Determine the complexity of $\hanano$ when the number of colored blocks is constant-bounded but the number of movable gray blocks is not. There is evidence that suggests that $\hanano$ is is not in $\NP$ in that case as the number of moves needed to solve such a level can be exponentially large~\cite{liu-yan:j:hanano}.
    \item It is unknown if $\jelly$ remains $\PSPACE$-complete when restricted to only two colors and no movable gray blocks, like $\hanano$. Crabtree and Mitsou~\cite{cra-mit:t:jelly} prove the result for a polynomial number of colors and no movable gray blocks. %
    \item While $\hanano$ and $\jelly$ are interesting in their own right,  it would be far more interesting (and perhaps very ambitious) to develop a more general framework, as has been done for games with an agent pushing or pulling blocks, \emph{\`a la} \cite{ani-bos-dem-dio-hen-lyn:c:door-pspace-hard,ani-chu-dem-dio-hen-lyn:c:checked-gadgets,ani-dem-hen-lyn:j:in-out-gadgets}, in order to generalize the preliminary observations and results presented in this paper. Our approach to defining and using configuration graphs was inspired by those papers.
    \item The main frameworks generalizing games with pushing blocks typically assume the presence of an agent that moves blocks, and \cite{mit-hardness:t:push-1} provides a way of uniting such frameworks and the $\NCL$ framework. It would be interesting to explore how that unification can be used to study games like Hanano and Jelly\@.
\end{enumerate}

\begin{table*}[ht]

\centering
\small
\begin{talltblr}
    [
    theme= CaptionNotIndented,
    caption = {A summary of the known complexity results about $\hanano$ and $\jelly$ when the restrictions are about the number of colors, the presence of movable gray blocks, and the bound of the number of colored blocks. 
    Any result attributed to a theorem in this paper is due to us.%
    },
    label = {table:summary},
    note{$\dagger$} = 
    {The construction given by Liu and Yang~\cite[Figure~12]{liu-yan:j:hanano} is quite insightful; they provide a level of Hanano with a single colored block and multiple movable gray blocks that generalizes to show that the number of moves under the constraints of the current row the table is exponential in the size of the level. Thus we suspect that the problem remains \PSPACE-complete when there are movable gray blocks but only a constant number of colored blocks. The methods we use are however not well-suited to provide \PSPACE-hardness reductions using only a constant number of colored blocks, and we suspect a new strategy would need to be devised to provide such reductions. 
    That same construction by~\cite{liu-yan:j:hanano} can easily be modified to hold analogously for $\jelly$.}
    ]
    {
    colspec = {X[1.7cm]|X[1.9cm]||X[1.2cm]|X|X[1.8cm]|X[31mm]},
    row{even} = {c,gray!25},
    hlines,
    row{1-2} = {font=\bfseries,gray!50},
    row{odd} = {c},
    rowsep= {1pt},
    colsep = {3pt}
    }
    \SetCell[r=2]{c} Movable Gray Blocks? & \SetCell[r=2]{c} \# of Colored Blocks & \SetCell[c=2]{c} {\sc Hanano} & & \SetCell[c=2]{c} {\sc Jelly} & \\
    &  & Colors & Result & Colors & Result\\ \hline
     Yes & Constant &
     1 & exp. moves\TblrNote{$\dagger$}& 
     1 & exp. moves\TblrNote{$\dagger$}\\
     Yes & Unbounded &
     1 & $\PSPACE$-c, \cite{cha:j:hanano} &
     1 & $\PSPACE$-c, \cite{cra-mit:t:jelly}\\
     No & Constant & 
     1 & $\P$, Thm.~\ref{theorem:hanano-constant-movable} & 
     1 & $\P$, Thm.~\ref{theorem:jelly-constant-movable}\\
     No & Unbounded & 
     1 & $\PSPACE$-c, Thm.~\ref{theorem:pspace-completeness} & 
     1 & $\NP$-c, \cite{yan:c:jelly}\\
     Yes & Constant & 
     $\geq 2$ & exp. moves\TblrNote{$\dagger$}& 
     Unbounded & exp. moves\TblrNote{$\dagger$}\\
     Yes & Unbounded & 
     $\geq 2$ & $\PSPACE$-c, \cite{cha:j:hanano} &
     Unbounded & $\PSPACE$-c, \cite{cra-mit:t:jelly}\\
     No & Constant & 
     $\geq 2$ & $\P$, Thm.~\ref{theorem:hanano-constant-movable} &
     Unbounded & $\P$, Thm.~\ref{theorem:jelly-constant-movable}\\
     No & Unbounded & 
     $\geq 2$ & $\PSPACE$-c, \cite{cha:j:hanano} &
     Unbounded & $\PSPACE$-c, \cite{cra-mit:t:jelly}\\
\end{talltblr}
\end{table*}

\clearpage
\bibliographystyle{alpha}
\newcommand{\etalchar}[1]{$^{#1}$}

\clearpage

\normalsize
\appendix

\newcommand{\gameplayshotsizes}{0.3\linewidth}

    \section{Solving Level~1 of Hanano}\label{app:solve-hanano-level1}
    We here provide the full set of steps used to solve the first level of Hanano to help with the reader's understanding of the mechanisms of the game.

    \begin{figure}[ht]
        \centering
        \begin{subfigure}[c]{\gameplayshotsizes}
            \includegraphics[width=\linewidth]{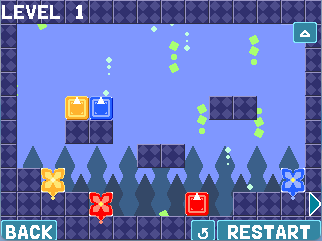}
            \caption{The initial configuration of Level~1.\\\\}
        \end{subfigure}
        \hfill
        \begin{subfigure}[c]{\gameplayshotsizes}
            \includegraphics[width=\linewidth]{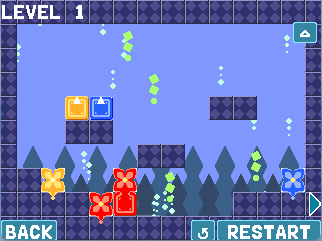}
            \caption{The red block moves left and blooms.\\\\}
        \end{subfigure}
        \hfill
        \begin{subfigure}[c]{\gameplayshotsizes}
            \includegraphics[width=\linewidth]{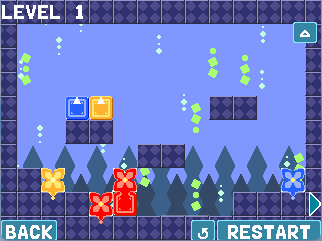}
            \caption{The blue and yellow blocks swap positions in one step.\\}
        \end{subfigure}

        \begin{subfigure}[c]{\gameplayshotsizes}
            \includegraphics[width=\linewidth]{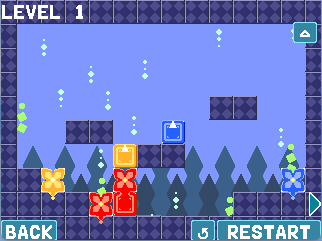}
            \caption{The yellow block moves right, falls, and allows the blue block to move further right.}
        \end{subfigure}
        \hfill
        \begin{subfigure}[c]{\gameplayshotsizes}
            \includegraphics[width=\linewidth]{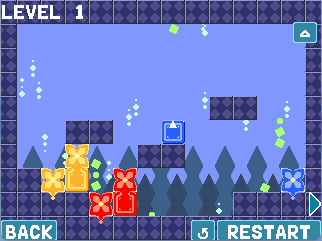}
            \caption{The yellow block then moves to the left and blooms by making contact with the yellow flower.}
        \end{subfigure}
        \hfill
        \begin{subfigure}[c]{\gameplayshotsizes}
            \includegraphics[width=\linewidth]{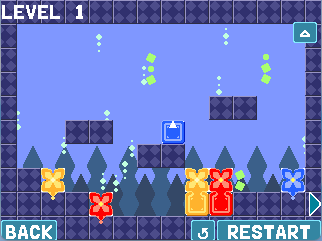}
            \caption{The (bloomed) red and yellow blocks are moved as far right as possible.\\}
        \end{subfigure}

        \begin{subfigure}[c]{\gameplayshotsizes}
            \includegraphics[width=\linewidth]{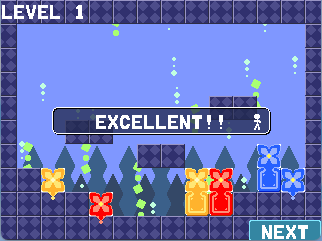}
            \caption{The blue block slides on top of the bloomed blocks to reach the blue flower.}
        \end{subfigure}        
        \caption{Steps to solve Level~1 of Hanano.}
        \label{fig:solve-hanano-level1}
    \end{figure}

\end{document}